\documentclass{sig-alternate-05-2015}

\usepackage{amssymb}
\usepackage{amsmath}

\usepackage{algorithm,algorithmic}

\usepackage{float}

\usepackage{subcaption}
\usepackage[table]{xcolor}
\usepackage{graphicx}
\usepackage{url}
\usepackage{graphics}
\usepackage{booktabs}
\usepackage{multirow}
\usepackage{dcolumn}
\usepackage{color}
\newlength{\thinline}
\newlength{\thickline}
\usepackage{enumitem}

\newcommand{\usr}{u}
\newcommand{\Usr}{\mathcal{U}}
\newcommand{\loc}{\ell}
\newcommand{\Loc}{\mathcal{L}}

\newcommand{\LocHour}{\mathcal{Q}}
\newcommand{\lochour}{q}

\newcommand{\UsrHour}{\mathcal{Y}}
\newcommand{\usrhour}{y}

\newcommand{\cathour}{ct}
\newcommand{\cat}{c}

\newcommand{\G}{\mathcal{G}}
\newcommand{\W}{\mathcal{W}}
\newcommand{\w}{w}

\newcommand{\Context}{\mathcal{N}}
\newcommand{\context}{n}

\newcommand{\feature}[1]{\varphi_{#1}}

\newcommand{\cicnt}[1]{\tau_{#1}}

\newcommand{\bias}[1]{\kappa_{#1}}

\begin{document}

\setcopyright{acmcopyright}





%

\title{DeepCity: A Feature Learning Framework for Mining Location Check-ins}

\numberofauthors{2}
\author{
\alignauthor
  Jun Pang\\
  \affaddr{FSTC \& SnT}\\
  \affaddr{University of Luxembourg}\\
  \email{jun.pang@uni.lu}
\alignauthor
  Yang Zhang\\
  \affaddr{FSTC}\\
  \affaddr{University of Luxembourg}\\
  \email{yang.zhang@uni.lu}
}
\maketitle
\begin{abstract}
Online social networks being extended
to geographical space has resulted in 
large amount of user check-in data.
Understanding check-ins 
can help to build appealing applications,
such as location recommendation.
In this paper,
we propose DeepCity,
a feature learning framework based on deep learning,
to profile users and locations, 
with respect to user demographic and location category prediction.
Both of the predictions 
are essential for social network companies to increase user engagement.
The key contribution of DeepCity
is the proposal of task-specific random walk
which uses the location and user properties
to guide the feature learning to be specific to each prediction task.
Experiments conducted on 42M check-ins
in three cities collected from Instagram 
have shown that DeepCity achieves a superior performance 
and outperforms other baseline models significantly.
\end{abstract}




\section{Introduction}
\label{sec:introduction}

The advancement of ICT technologies has extended
online social networks (OSNs) to geographical space.
Nowadays, in almost all popular social network platforms,
including Facebook, Twitter and Instagram,
users frequently sharing their photos or statues
together with geographical locations, namely \emph{check-ins}.
The large quantity of user check-in data
has provided us with an unprecedented chance 
to study users' mobility behaviors.
Understanding it can result in many appealing applications,
such as location recommendation.
In addition, it can also help us to tackle some of the crucial problems
we are facing at the moment, 
including air pollution control and epidemiology study.

One important aim of understanding user check-ins
is to profile users
with the assumption that
whereabouts of a user 
reflects who he is.
Profiling users is essential for OSNs
since it can help to increase user engagement.
For instance, 
by knowing a user is female, 
an OSN company
can recommend her to a certain place
where other female users like to visit as well.
Alternatively, as pointed out by~\cite{DHVB13},
mobility is among the most 
sensitive information being collected 
from each individual in the Internet era.
Knowing to which extent a user's personal information can be inferred
based on his mobility
can provide in depth guidances on how to design systems to 
protect user privacy.

On the other side of the coin,
check-in data can also help us
to gain more understandings about locations,
i.e., location profiling.
One important problem in this direction
is using users' check-in behavior at a certain location
to infer the location's category.
Since most of the location information
in OSNs is crowdsourced by users
and many information is incomplete,
begin able to infer a location's category can be particularly useful
since the result 
can facilitate people to explore new areas,
increase the performance of location recommender and so on.

Mining user check-ins 
has attracted academia a considerable amount of attention.
Researchers have tackled various problems,
including the above discussed user and location profiling,
with the help of machine learning,
e.g.,~\cite{CML11,SNM11,WLL14}.
However, these tasks, most of which involve prediction,
require hand-engineering domain-specific features
for learning algorithms.
Except for the tedious efforts spent on feature engineering,
these features, in many cases, are not complete.

The recent development of deep learning~\cite{MCCD13,MSCCD13}
has provided an alternative way 
to automatically learn features
for prediction tasks based on network structures~\cite{PAS14,TQWZYM15,GL16}.
In this setting,
features are learned by optimizing a general objective function,
and thus can be applied in any prediction tasks.
These methods are also referred as network embedding,
and the learned features of each node
is named as the node's embedded vector.
However, we notice that these network embedding approaches
often ignore the specific knowledge
with respect to different prediction tasks, 
which eventually affects the prediction's performance.

In the current paper,
we propose DeepCity,
a general feature learning framework
for mining user check-ins shared in OSNs.
We concentrate on two data mining tasks:
user profiling, represented by demographic inference,
and location profiling, represented by location category inference,
both of which are essential for OSNs
as we discussed previously.

DeepCity adopts the state-of-the-art
network embedding method, namely the Skip-gram model,
based on deep learning techniques.
With an object function
aiming to preserve each user's or location's neighbor information,
features to be fed into learning algorithms for prediction tasks 
are automatically computed.
Therefore, hand-engineering features are no longer necessary.

DeepCity's key contribution 
is the proposal of task-specific random walk:
for each prediction,
task-specific random walk utilizes location or user properties
with respect to the prediction
to guide the algorithm to define each user's (location's) 
neighbors to be more specific to the prediction.
In this way, 
the learned features capture the useful information for the prediction.
Our experiments have been conducted on a large-scale check-in dataset
collected in three major cities in the world, 
including New York, Los Angeles and London, from Instagram.
Extensive experiments have shown
that DeepCity 
has outperformed the existing state-of-the-art models significantly.

In summary, we make the following contributions:
\begin{itemize}[leftmargin=*]
 \itemsep-0.5mm
 \item We propose a feature learning framework based on deep learning,
 namely DeepCity, for predicting user demographics 
 and location category.
 We propose task-specific random walk,
 which allows DeepCity to learn different features for different predictions.
 DeepCity takes the advantage of efficiency 
 raised by the Skip-gram model
 over other feature learning methods,
 and it achieves superior performances over general network embedding models.
 \item We construct a large dataset
 with more than 40M check-ins in three cities in the world.
 Each check-in is affiliated with the location's detailed information,
 such as name and category.
 In addition, we utilize a state-of-the-art facial recognition tool, namely Face++,
 to collect users' demographic information at a large scale.
 The dataset provides us with resourceful information
 to conduct various user check-in mining tasks in the future.
 (The dataset and experimental code are available upon request.)
 \item We conduct extensive experiments with DeepCity using the Instagram dataset.
 Evaluation results have demonstrated our framework's 
 superior performances over other baselines.
 In addition to this, we further study the parameter sensitivity
 and the robustness on how to choose active users,
 in terms of the number of check-ins per user, for experiments.
 Experimental results show that 
 even when concentrating only 
 on users with at least 10 check-ins,
 DeepCity still achieves a strong performances, 
 which further demonstrates our framework's robustness.
\end{itemize}

The rest of the paper is organized as the following.
The DeepCity framework is described in Section~\ref{sec:framework}.
We present the experimental results in Section~\ref{sec:experiments}.
Section~\ref{sec:relwork} introduces the related works.
In Section~\ref{sec:conclu}, we conclude the paper and 
present some interesting future directions.

\section{DeepCity Framework}
\label{sec:framework}

In this section, we introduce our DeepCity framework.
We start by describing the notations used throughout the paper,
then introduce the feature learning algorithm utilized in DeepCity, 
i.e., the Skip-gram model.
In the end, we present task-specific random walk
which leads to user and location feature learning for prediction.

\subsection{Notations}

We denote each user by $\usr$
and each location by $\loc$,
two sets $\Usr$ and $\Loc$ contain all the users and locations, respectively.
For demographic prediction,
instead of only considering a user's spatial information,
we take one step further to consider both his spatial and temporal information,
thus we introduce another notation $\lochour\in \LocHour$ 
to denote a temporal-location,
it is used to describe a user visits \emph{a certain location at certain time}.
In the current paper,
the temporal information is considered in a 24-hour scale,
thus $\LocHour\subseteq \Loc\times \{1, 2, \ldots, 24\}$.
Similarly, for location category prediction,
we define a notion namely temporal-user, denoted by $\usrhour$,
which is used to describe a location is visited by \emph{a certain user at certain time},
$\UsrHour\subseteq \Usr\times\{1, 2, \ldots, 24\}$ contains all the temporal-users.
We denote the bipartite graph between $\Usr$ and $\LocHour$
by $\G_{\Usr,\LocHour} = (\Usr, \LocHour, \W_{\Usr, \LocHour})$
where $\W_{\Usr, \LocHour}$ represents the weighted edges between users and temporal-locations.
Correspondingly,
$\G_{\Loc,\UsrHour} = (\Loc, \UsrHour, \W_{\Loc, \UsrHour})$
represents the location temporal-user graph.
We define the edge weight
from $\usr$ to $\lochour$, and
from $\lochour$ to $\usr$ in $\G_{\Usr,\LocHour}$ as
\[
 \w_{\usr, \lochour} = \frac{\cicnt{\usr, \lochour}}{\cicnt{\usr}}\mbox{  and  }
 \w_{\lochour, \usr} = \frac{\cicnt{\usr, \lochour}}{\cicnt{\lochour}},
\]
respectively.
The weights reflect how frequent $\usr$ visits $\lochour$ and
$\lochour$ is visited by $\usr$.
Similarly, the edge weight from $\loc$ to $\usrhour$,
and from  $\usrhour$ to $\loc$
in $\G_{\Loc, \UsrHour}$ are defined as
\[
 \w_{\loc, \usrhour} = \frac{\cicnt{\loc, \usrhour}}{\cicnt{\loc}} \mbox{  and }
 \w_{\usrhour, \loc} = \frac{\cicnt{\loc, \usrhour}}{\cicnt{\usrhour}}.
\]
Here, $\cicnt{\usr,\lochour}$ ($\cicnt{\loc,\usrhour}$)
represents $\usr$'s ($\loc$'s) 
number of check-ins at $\lochour$ (by $\usrhour$),
while $\cicnt{\usr}$ and $\cicnt{\lochour}$ 
($\cicnt{\loc}$ and $\cicnt{\usrhour}$)
represent $\usr$'s and $\lochour$'s ($\loc$'s and $\usrhour$'s) 
total number of check-ins, respectively.

\subsection{Skip-gram model}

The Skip-gram model proposed in~\cite{MCCD13,MSCCD13}
is first designed to embed words in documents
into a continuous vector space (word2vec),
with the object that a word's embedded vector
can predict its nearby words.
Alternatively, 
Skip-gram preserves a word's nearby words' information in the word's vector.
Skip-gram can be considered as a (shallow) neural network with one hidden layer,
and the learned word vectors, treated as features,
are demonstrated to be effective
on solving various sophisticated natural language processing (NLP) tasks,
such as document comparison~\cite{KSKW15}.

Perrozi et al.~\cite{PAS14} take the Skip-gram model
into the network mining field, namely DeepWalk,
by establishing an analogy that 
a node in a  network is a word
and the network itself is a document.
DeepWalk creates ``sentences'' 
by simulating random walk traces in the network
and feed them into the Skip-gram model
to obtain the vector representation of each node.
Similar to its assumption in NLP,
the Skip-gram model in networks 
preserves the information of each node's nearby nodes
in random walk traces (neighbor nodes),
and the learned features have been applied in multiple data mining tasks
to achieve superior performances,
such as multi-label classification.
DeepWalk also inspires many recent state-of-the-art network embedding methods,
including LINE~\cite{TQWZYM15} and node2vec~\cite{GL16}.

Based on the previous works,
we introduce the Skip-gram model
into user check-in mining.
For simplicity,
we only present the model for $\G_{\Usr, \LocHour}$,
and the model for $\G_{\Loc, \UsrHour}$ can be deduced easily.
The Skip-gram model is formalized into 
a maximal likelihood optimization problem,
with the following object function:
\begin{equation}\label{equ:gen}
  \underset{\theta}{\mathrm{argmax}}\prod_{v\in \Usr\cup\LocHour} p(\Context(v) \vert v; \theta).
\end{equation}
Here, $\Context(v)$ represents the neighbor nodes of $v$ 
and $\theta$ represents the parameters of the model.
By assuming that neighbor nodes prediction
is independent of each other,
Equation~\ref{equ:gen} is factorized into
\begin{equation}
  \underset{\theta}{\mathrm{argmax}}\prod_{v\in \Usr\cup\LocHour}\prod_{\context \in \Context(v)} p(n \vert v;\theta).
\end{equation}
To model the conditional probability $p(n\vert v;\theta)$,
a softmax function is adopted:
\begin{equation}
  p(n\vert v;\theta) = \frac{e^{\feature{\context}\cdot \feature{v}}}
  {\underset{{m\in \Usr\cup\LocHour}}{\sum} e^{\feature{m}\cdot \feature{v}}}
\end{equation}
where $\feature{v}$ and $\feature{\context}$ 
represent the vectors for node $v$ and its neighbor $\context$, respectively,
and $\feature{\context}\cdot \feature{v}$ is the two vectors' dot product.
Both $\feature{v}$ and $\feature{\context}$ belong to the parameters $\theta$,
and we set the vector length to $d$ for all vectors.
In the end, by adding every piece together
and switching the multiplication to summation through log-likelihood,
we obtain the following optimization object:
\begin{equation}\label{equ:final}
  \underset{\theta}{\mathrm{argmax}}\!\!\sum_{v\in \Usr\cup\LocHour}
  \sum_{\context \in \Context(v)} (\feature{\context}\! \cdot\! \feature{v}  -  
  \log(\!\!\underset{{m\in \Usr\cup\LocHour}}{\sum}\!\!e^{\feature{m}\cdot \feature{v}})).  
\end{equation}
The term $\sum_{m\in \Usr\cup\LocHour} e^{\feature{m}\cdot \feature{v}}$
is expensive to compute, especially for large networks,
since it requires the summation over all nodes,
thus negative sampling is proposed~\cite{MSCCD13}
to improve Skip-gram's learning efficiency,
which we adopt in the current paper as well.
In the end,
we utilize stochastic gradient decent (SGD)
to optimize the object function
to obtain the vector of each node $v$, i.e., $\feature{v}$.

It is worth noticing that the previous network embedding models,
including DeepWalk and node2vec,
mainly concentrate on homogeneous networks,
such as social networks,
while, in our setting, we extend it to bipartite graphs, 
i.e., $\G_{\Usr, \LocHour}$ and $\G_{\Loc, \UsrHour}$.
This indicates that a user $\usr$'s neighbors
is composed by both temporal-locations and users.
The same applies for a location $\loc$.
This extension is straightforward and
preserves the original structure of user check-in data
without modifying it, such as using 2-hop edges 
to construct a homogeneous user network,
therefore, we keep the model in this way.
As we shall see in Section~\ref{sec:experiments}, 
DeepCity achieves a superior performance 
in demographic and location category prediction,
which indicates that the Skip-gram model 
can be directly applied on bipartite graphs.

\subsection{Task-specific random walk}

\noindent\textbf{Intuition.}
The Skip-gram model aims to preserve the neighbors of each node.
To define a node's neighbors,
DeepWalk and node2vec
adopt the random walk approach,
i.e., a node's neighbors are those 
that are close to it in random walk traces.
The learned vectors under these models
are not specific to any prediction problems,
which make them applicable to any tasks.
However, for each individual prediction task, 
random walk approach does not fully take into account 
useful information with respect to the prediction,
which may affect Skip-gram's performances.
To give an example,
if several users frequently visit the same metro station
and office at the same hour,
then they should be close to each other 
in the learned vector spaces since they have similar neighbors
in random walk traces (composed by temporal-locations and users),
which will result in them being partitioned
into the same class by machine learning classifiers.
However, the neighbors introduced by the above mentioned locations
does not provide much information about 
a user's demographics:
they may use the same metro to commute
and go to office to work.
On the other hand, if we know that these users often
go to locations that are ``biased'' towards 
people of certain demographics,
then we are more confident
that they belong to the same demographic group.
Here, ``bias'' means that
a temporal-location's (temporal-user's) check-in distribution
over a certain demographic (location category)
is quite different from the general check-in distribution 
over the demographic (location category).

To take into account the bias 
for specific prediction tasks,
we propose task-specific random walk.
Our intuition is that:
\emph{
if the neighbors of $\usr$ ($\loc$)
generated by random walks
contains useful information on $\usr$'s demographics ($\loc$'s category),
then the features learned by Skip-gram,
which preserves the neighbor information of $\usr$ ($\loc$),
can be used to effectively predict $\usr$'s demographics ($\loc$'s category),
with learning algorithms.}

\smallskip
\noindent\textbf{Algorithm.}
To proceed,
we first propose a measurement
to quantify to which extent 
a temporal-location (temporal-user)
is biased towards a certain demographic group (location category).
As discussed above,
the bias refers to the distribution difference,
to capture it,
we resort to Kullback-Leibler divergence (KL divergence),
a classical approach for comparing distributions.
Taking gender as an example,
we denote the number of all check-ins made by female users
as $\cicnt{{\it female}}$ (male users as $\cicnt{{\it male}}$),
and use $p({\it female}) = \frac{\cicnt{{\it female}}}{\cicnt{{\it female}} + \cicnt{{\it male}}}$
($p({\it male}) = \frac{\cicnt{{\it male}}}{\cicnt{{\it female}} + \cicnt{{\it male}}}$)
to represent the check-in proportion of female users (male users).
For a temporal-location $\lochour$,
we use $\cicnt{\lochour, {\it female}}$ ($\cicnt{\lochour, {\it male}}$) 
to denote the number of check-ins made by female (male) users at $\lochour$,
and define the proportion
$p(\lochour, {\it female}) = 
\frac{\cicnt{\lochour, {\it female}}}{\cicnt{\lochour, {\it female}} + \cicnt{\lochour, {\it male}}}$
($p(\lochour, {\it male}) = 
\frac{\cicnt{\lochour, {\it male}}}{\cicnt{\lochour, {\it female}} + \cicnt{\lochour, {\it male}}}$).
Then,
the gender biased value of $\lochour$ is defined as
\[
  \bias{\lochour, {\it gender}} = 
  \underset{i \in \{{\it female},{\it male}\}}{\sum}
  p(\lochour, i)  \log\frac{p(\lochour, i)}{p(i)}.
\]
Our gender biased value follows 
the original definition of KL divergence,
higher value indicates larger difference between two distributions.
Table~\ref{table:gender_kl} 
depicts the top 10 gender biased temporal-locations
in three cities in our datasets (Section~\ref{sec:experiments}).
In New York and Los Angeles, 
the most male biased temporal-locations are gyms in the morning,
such as FIT RxN (7h) and 24 Hour Fitness (8h).
In London, football stadiums in the evening 
are mainly men's places:
Emirates Stadium and Stamford Bridge are home courts
for two famous football clubs in London.
Meanwhile, the most female biased temporal-locations
include library (Jefferson Market Library), performing art place (Sherry Theatre)
and theme restaurant (sketch).
The biased value of the other two demographics studied
in the current paper, i.e., race and age (see Section~\ref{sec:experiments}),
is defined accordingly.
We also utilize KL divergence
to quantify the location category biased value of each temporal-user.

\begin{table*}[!t]
  \centering
  \begin{tabular}{c|c|c}
  \toprule[\thickline]
  New York & Los Angeles & London \\
  \midrule[\thinline]
  FIT RxN (7h) & 24 Hour Fitness (8h) & Emirates Stadium (12h) \\
  FIT RxN (8h) & 24 Hour Fitness (9h) & \cellcolor{gray!25} Sea Life London Aquarium (10h) \\
  FIT RxN (10h) & LA Fitness (19h) & Stamford Bridge (17h) \\
  5050 Skatepark (23h) &  Micky's (1h) & Emirates Stadium  (22h) \\
  The James Beard House (19h) & \cellcolor{gray!25} Sherry Theatre (8h) & Stamford Bridge (19h) \\
  Alaska (2h) & \cellcolor{gray!25} Sherry Theatre (9h)& \cellcolor{gray!25} Horniman Museum and Gardens (13h) \\
  The James Beard House (20h) & Micky's (23h) & Stamford Bridge (22h) \\
  Stage 48 (11h) & The Satellite (21h) & \cellcolor{gray!25} sketch (16h) \\
  \cellcolor{gray!25} Jefferson Market Library (7h) & The Abbey (1h) & Cafe Oto (21h) \\
  \cellcolor{gray!25} P.S.\ 41 (7h) & The Abbey (19h) & \cellcolor{gray!25} sketch (22h) \\
  \bottomrule[\thickline]
  \end{tabular}
  \caption{The most gender biased locations, gray boxes mark the female biased locations.\label{table:gender_kl}}
\end{table*}

After obtaining the biased 
values of each temporal-location
with respect to all demographics,
we extend $\G_{\Usr, \LocHour}$
into $\G_{\Usr, \LocHour}^{\rho} = (\Usr, \LocHour, \W_{\Usr, \LocHour}^\rho)$
where ${\rho} \in \{{\it gender}, {\it race}, {\it age}\}$,
i.e., $\G_{\Usr, \LocHour}$ is extended into three instances.
The weight between $\usr$ and $\lochour$ in the new extended 
$\G_{\Usr, \LocHour}^\rho$ is defined as
\[
 \w_{\usr,\lochour}^\rho = \frac{\cicnt{\usr, \lochour}}{\cicnt{\usr}}  \bias{\lochour, \rho}
\]
where $\bias{\lochour, \rho}$ represents 
the $\rho$ biased value
of the temporal-location $\lochour$.
On the other hand, the edge from $\lochour$ to $\usr$ 
in $\G_{\Usr, \LocHour}$ stays unchanged,
i.e., $\w_{\lochour, \usr}^\rho = \w_{\lochour, \usr}$.
This way of edge weight definition 
drives our random walks to be biased 
towards $\rho$ biased temporal-locations,
and users frequently visiting these $\rho$ biased locations,
which in the end results in each user's neighbors 
being able to reflect his demographic information,
i.e., the intuition of task-specific random walk.
Meanwhile, $\G_{\Loc, \UsrHour}$ is extended into 
$\G_{\Loc, \UsrHour}^{\varrho} = (\Loc, \UsrHour, \W_{\Loc, \UsrHour}^\varrho)$
based on the location category biased value of each temporal-user,
and edge weights in $\G_{\Loc, \UsrHour}^\varrho$
is modified accordingly.

The task-specific random walk
is the normal random walk (on weighted graphs)
executed on the above introduced extended bipartite graphs.
The walk traces obtained are feed into Skip-gram for embedding.
Algorithm~\ref{alg:gen} presents DeepCity
for demographic prediction.
The algorithm starts the random walk
from each node in $\G_{\Usr,\LocHour}^\rho$,
and repeats it for $r$ times, i.e., the number of walks per node.
Each walk takes $s$ steps long, i.e., the length of each walk.
The task-specific random walk is presented in Algorithm~\ref{alg:rw},
for each current node, 
we extract its neighbors and the corresponding weights,
and utilize a sampling algorithm that takes into account weights
to find the next node.
The generated random walk traces are 
then fed into a SGD optimizer,
to learn vectors of all nodes.

\begin{algorithm}[!h]
\caption{DeepCity for demographic $\rho$\label{alg:gen}}
\begin{algorithmic}[1]
\REQUIRE{$\G_{\Usr,\LocHour}^{\rho} = (\Usr, \LocHour, \W_{\Usr,\LocHour}^{\rho} )$, 
walk length $s$, number of walks per node $r$, 
vector dimension $d$}
\STATE{${\it walks} \leftarrow [ \ ]$}
\FOR{i = 1 to $r$}
  \FOR{$v \in \Usr\cup \LocHour$}
    \STATE{${\it v\_ walk \leftarrow}$ TRW$(\G_{\Usr,\LocHour}^{\rho}, v, s)$}
    \STATE{$\mbox{append } {\it v\_ walk} \mbox{ to }{\it walks}$}
  \ENDFOR
\ENDFOR
\STATE{$\theta\leftarrow \mbox{SGD}({\it walks}, d)$}
\RETURN{$\theta$}
\end{algorithmic}
\end{algorithm}

\begin{algorithm}[!h]	
\caption{TRW (task-specific random walk)\label{alg:rw}}
\begin{algorithmic}[1]
\REQUIRE{$\G_{\Usr,\LocHour}^{\rho}, v, s$}
\STATE{${\it v\_ walk} \leftarrow [v]$}
\STATE{${\it curr} \leftarrow v$ \# initialize the current node}
\FOR{i = 2 to $s$}
  \STATE{${\it curr\_ nb},{\it curr\_ nb\_w} \leftarrow \mbox{GetNeighbour}({\it curr}, \G_{\Usr,\LocHour})$\\ 
  \# obtain {\it curr}'s neighbors and the weights between {\it curr} and its neighbors}
  \STATE{${\it next} \leftarrow \mbox{SamplingWithWeight}({\it curr\_ nb},{\it curr\_ nb\_w})$}
  \STATE{$\mbox{append } {\it next} \mbox{ to }{\it u\_walk}$}
  \STATE{${\it curr} \leftarrow {\it next}$}
\ENDFOR
\RETURN{${\it v\_ walk}$}
\end{algorithmic}
\end{algorithm}

The learned vectors are treated as features of each node,
and are directly feed into machine learning classifiers,
in order to obtain 
the final prediction results\footnote{
We simply omit the features of those nodes
that are not the target of each prediction task,
i.e., temporal-locations in demographic prediction 
and temporal-users in location category prediction.}.

\smallskip
\noindent\textbf{Advantages.}
There are mainly two advantages for our approach.
The first one is efficiency.
As pointed out by~\cite{GL16},
one of the advantages of Skip-gram based network embedding method
is its efficiency 
in terms of both time and space requirements
over other feature learning methods,
such as matrix factorization.
Since our approach also follows Skip-gram,
then it is more efficient over other methods.
However, as DeepCity
requires simulating random walks for 
each individual prediction task,
thus it is less efficient than those general approaches,
such as DeepWalk and node2vec,
whose vectors can be applied to any prediction tasks.
However, the computing overhead of our approach 
simply depends on the number of prediction tasks we are given
which are naively parallelizable,
thus this should not be a major concern.
Moreover, as we define different random walks 
for different prediction tasks,
the performance of DeepCity should be stronger 
than the general random walk approaches,
which leads to our framework's second advantage:
effectiveness,
which we will demonstrate in Section~\ref{sec:experiments}.


\section{Experiments}
\label{sec:experiments}

In this section, we first introduce the Instagram dataset 
we have collected for conducting experiments.
Next, we describe our experiment setup 
including baseline models and evaluation metrics.
In the end, we present the experimental results
together with parameter sensitivity study 
and robustness study.

\subsection{Datasets}
Instagram is the second largest OSN,
and it leads the market of photo sharing.
According to a study~\cite{MHK14},
Instagram users are more likely to share their locations
than other social networks', such as Twitter,
which makes Instagram a suitable platform for collecting user check-ins.

\begin{figure}[!t]
  \centering
  \includegraphics[width=0.7\columnwidth]{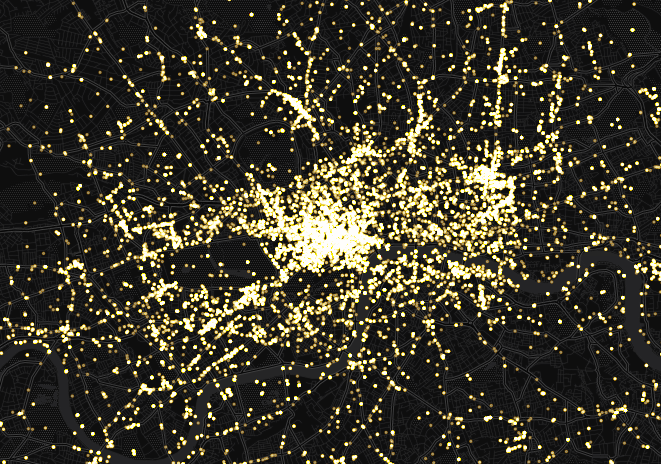}
  \caption{A sample of check-ins in London.\label{fig:london_checkin}}
\end{figure} 

Our datasets are collected through Instagram's REST API.
We concentrate on three cities in the world,
including New York, Los Angeles and London.
To collect Instagram users' check-ins,
we first resort to the API of Foursquare,
a popular location-based social network
whose API is linked 
with Instagram's\footnote{Connection aborted on 2016
(\url{http://bit.ly/1Ne85wO}).},
to collect Foursquare's location ids in the three cities.
In this way, we are able to get not only users' check-ins,
but also the detailed location information of these check-ins,
such as location name and location category,
the latter is used as labels for location category prediction.
Foursqaure organizes location categories 
into a tree structure\footnote{\url{http://bit.ly/2dJDgns}},
we adopt the first level categories in the tree,
including Entertainment, University, Food, Nightclub, Outdoor, Professional,
Residence, Shop and Transportation,
as labels for location category prediction.
Next, we query Instagram's API 
with previously collected Foursquare's location ids
to get the corresponding Instagram's location ids,
which in the end leads to all check-ins of each location
since the beginning when the location is created in Instagram's platform.
Each check-in is organized into the following form.
\[
 \langle {\it user\ id}, {\it time}, {\it lattitude}, {\it longitude}, {\it location\ id}\rangle
\]

To obtain each users' demographic information,
we resort to Face++\footnote{\url{http://www.faceplusplus.com/}}, 
a state-of-the-art facial recognition
software based on deep neural networks,
to analyze users' profile photos.
Face++ takes a photo as input and returns 
gender, race (Asian, White, African) 
and age of the user (or users) in the photo.
Besides, Face++ also provides confidence scores
for its gender and race recognition, 
and age range for age recognition.
It is worth noticing that Face++ has been used in many works
to obtain users' demographics for analysis, 
e.g., \cite{SCFYCQA15,RQGG15,MLR15}.
To increase the data quality for our experiments,
we only concentrate on the data 
from users with more than 95\% gender and race confidence,
meanwhile, 
only users whose age range is equal or less than 5 are taken into account.
In addition, users whose profile photos contain more than one person
are filtered out.
Following previous works, such as~\cite{DYTYC14},
we model age prediction as a classification problem.
To define age labels, we adopt the methods in~\cite{FSPLd2015}
to discretize age into three equal groups:
15-20, 21-25 and 26-36.
There are two reasons that our users are at a younger age.
First, Instagram 
is popular among younger people\footnote{\url{http://pewrsr.ch/1foZCGR}}.
Second, the previous data processing 
(95\% gender and race confidence and age range $\leq 5$)
filters out most of the users aged older than 36.

In the end, we collect more than 19.6M check-ins in New York,
14.9M in Los Angeles and 8.4M in London.
Figure~\ref{fig:london_checkin}
depicts a sample of check-ins collected in London.
To resolve the data sparseness issue,
we further concentrate on users with at least 20 check-ins,
whom we term as active users.
Later in our experiments, 
we also study the robustness of this number of check-ins
defining active users, with respect to our prediction tasks.
Table~\ref{table:dataset} summarizes our datasets.

\begin{table}[!h]
  \centering
  \begin{tabular}{ c|ccc}
  \toprule[\thickline]
  & New York & Los Angeles & London \\
  \midrule[\thinline]
  \#. check-ins & 19,600,207& 14,902,098 & 8,419,610 \\
  \#. users &1,165,286 & 900,737 & 823,694\\
  \#. locations & 54,961 & 50,681 & 25,659\\
  \#. active uses & 12,778 & 7,706 & 5,566\\
  \bottomrule[\thickline]
  \end{tabular}
  \caption{Dataset.\label{table:dataset}}
\end{table}

\subsection{Experiment setup}

We adopt the following baseline models for evaluating our demographic prediction:
\begin{itemize}[leftmargin=*]
  \itemsep-0.5mm
  \item DeepCity$(\loc)$: 
  We omit the temporal information (hour) considered 
  in DeepCity for demographic prediction, 
  i.e., the task-specific random walk is performed
  on a user location bipartite graph.
  \item DeepCity$(\cathour)$: 
  We replace fine-grained locations used in DeepCity 
  with location categories.
  The location categories here (381 categories) 
  are extracted from the second layer in Foursquare's category tree,
  different from the labels used in location category prediction.
  \item DeepCity$(\cat)$: 
  Similar to DeepCity$(\loc)$, this model omits
  the temporal information in DeepCity$(\cathour)$ model.
  \item STL~\cite{ZYZZX15}: 
  STL organizes users' check-ins into a three-way tensor:
  the first dimension represents users, 
  the second one represents spatial and temporal information,
  the third one represents location information, 
  including location category, reviews and keywords.
  A Tucker decomposition is applied on the tensor to transform users into vectors,
  based on which a support vector machine (SVM) is trained 
  for predicting users' demographics.
  Since our datasets do not contain reviews and keywords of each location,
  we only use category information for constructing the third dimension in STL.
  \item MF~\cite{KSG13}: 
  This model organizes user and location into a matrix, 
  and perform matrix factorization 
  to derive user vectors.
  Similar to STL, a SVM is adopted for prediction.
  \item DeepWalk~\cite{PAS14}: 
  This is a variant version of the original DeepWalk:
  random walk is performed 
  on $\G_{\Usr, \LocHour}$,
  i.e., no demographic biased value 
  are applied to adjust edge weights.
\end{itemize}

Both MF and DeepWalk generate 
not only user vector but also location vector,
which can be directly applied to infer location categories.
Besides, we also adopt the following baselines 
for evaluating location category prediction.
\begin{itemize}[leftmargin=*]
  \itemsep-0.5mm
  \item DeepCity$(\usr)$: 
  This model does not consider the temporal information 
  in DeepCity for predicting location category.
  \item Tensor: 
  Inspired by STL, 
  we organize location, user and hour into a three-way tensor
  and perform Tucker decomposition to obtain the latent vectors to represent locations.
  A SVM is trained for location category prediction.
  \item SAP~\cite{YSLYJ11}: SAP model takes into account two sets of features
  for inferring location category.
  The first set summarizes some explicit properties of each location, 
  such as its number of check-ins, number of users and visiting hours.
  The second set of features utilizes information retrieval techniques, 
  including random walk with restart and relaxation labeling,
  to find each location's neighbor locations' (through time and user) properties
  to infer the location's category.
  The two sets of features are combined together 
  to train a SVM to obtain the final prediction.
\end{itemize}

\begin{table*}[!ht]
  \centering
  \begin{tabular}{ c|cccc|cccc|cccc}
  \toprule[\thickline]
  & \multicolumn{4}{c|}{New York} & \multicolumn{4}{c|}{Los Angeles} & \multicolumn{4}{c}{London} \\
  \midrule[\thinline]
  & AUC & Pre. & Rec. & F1 & AUC & Pre. & Rec. & F1  & AUC & Pre. & Rec. & F1 \\
  \midrule[\thinline]
  \rowcolor{gray!25}
  DeepCity & 0.95 & 0.85 & 0.80 & 0.82 & 0.95 & 0.85 & 0.80 & 0.83 & 0.95 & 0.85 & 0.80 & 0.83 \\
  DeepCity$(\loc)$ & 0.82 & 0.72 & 0.47 & 0.57 & 0.82 & 0.72 & 0.49 & 0.58 & 0.83 & 0.73 & 0.52 & 0.60 \\
  DeepCity$(\cathour)$ & 0.71 & 0.60 & 0.29 & 0.39 & 0.73 & 0.61 & 0.32 & 0.43 & 0.71 & 0.62 & 0.34 & 0.43 \\
  DeepCity$(\cat)$ & 0.68 & 0.57 & 0.24 & 0.34 & 0.68 & 0.55 & 0.27 &0.36 & 0.68 & 0.58 & 0.29 & 0.38 \\
  STL & 0.65 & 0.60 & 0.13 & 0.22 & 0.65 & 0.58 & 0.13 & 0.21 & 0.67 & 0.61 & 0.16 & 0.25 \\
  MF & 0.58 & 0.48 & 0.06 & 0.10 & 0.61 & 0.51 & 0.08 & 0.15 & 0.59 & 0.46 & 0.07 & 0.13 \\
  DeepWalk & 0.68 & 0.57 & 0.23 & 0.33 & 0.65 & 0.53 & 0.22 & 0.31 & 0.67 & 0.56 & 0.25 & 0.35 \\
  \bottomrule[\thickline]
  \end{tabular}
  \caption{Experimental results on gender prediction.\label{table:predict_gender}}

  \bigskip
  
  \centering
  \begin{tabular}{c|cc|cc|cc}
  \toprule[\thickline]
  & \multicolumn{2}{c|}{New York} & \multicolumn{2}{c|}{Los Angeles} & \multicolumn{2}{c}{London} \\
  \midrule[\thinline]
  & Macro-F1 & Micro-F1 & Macro-F1 & Micro-F1 & Macro-F1 & Micro-F1 \\
  \midrule[\thinline]
  \rowcolor{gray!25}
  DeepCity & 0.72 & 0.93 & 0.66 & 0.92 & 0.62 & 0.95 \\
  DeepCity$(\loc)$ & 0.62 & 0.88 & 0.55 & 0.85 & 0.49 & 0.92 \\
  DeepCity$(\cathour)$ & 0.48 & 0.87 & 0.45 & 0.84 & 0.45 & 0.90 \\
  DeepCity$(\cat)$ & 0.43 & 0.86 & 0.43 & 0.83 & 0.40 & 0.90 \\
  STL & 0.33 & 0.85 & 0.35 & 0.83 & 0.32 & 0.90 \\
  MF & 0.31 & 0.85 & 0.31 & 0.82 & 0.32 & 0.91\\
  DeepWalk & 0.46 & 0.86 & 0.44 & 0.83 & 0.37 & 0.90 \\
  \bottomrule[\thickline]
  \end{tabular}
  \caption{Experimental results on race prediction.\label{table:predict_race}}

  \bigskip

  \centering
  \begin{tabular}{c|cc|cc|cc}
  \toprule[\thickline]
  & \multicolumn{2}{c|}{New York} & \multicolumn{2}{c|}{Los Angeles} & \multicolumn{2}{c}{London} \\
  \midrule[\thinline]
  & Macro-F1 & Micro-F1 & Macro-F1 & Micro-F1 & Macro-F1 & Micro-F1 \\
  \midrule[\thinline]
  \rowcolor{gray!25}
  DeepCity & 0.53 & 0.53 & 0.52 & 0.53 & 0.54 & 0.55 \\
  DeepCity$(\loc)$ & 0.40 & 0.41 & 0.41 & 0.42 & 0.45 & 0.45 \\
  DeepCity$(\cathour)$ & 0.38 & 0.38 & 0.38 & 0.39 & 0.39 & 0.40 \\
  DeepCity$(\cat)$ & 0.34 & 0.35 & 0.35 & 0.36 & 0.35 & 0.36 \\
  STL & 0.35 & 0.35 & 0.34 & 0.35 & 0.33 & 0.35\\
  MF & 0.34 & 0.34 & 0.33 & 0.35 & 0.32 & 0.35\\
  DeepWalk & 0.35 & 0.36 & 0.35 & 0.36 & 0.34 & 0.36 \\
  \bottomrule[\thickline]
  \end{tabular}
  \caption{Experimental results on age prediction.\label{table:predict_age}}

  \bigskip
  
  \centering
  \begin{tabular}{c|cc|cc|cc}
  \toprule[\thickline]
  & \multicolumn{2}{c|}{New York} & \multicolumn{2}{c|}{Los Angeles} & \multicolumn{2}{c}{London} \\
  \midrule[\thinline]
  & Macro-F1 & Micro-F1 & Macro-F1 & Micro-F1 & Macro-F1 & Micro-F1 \\
  \midrule[\thinline]
  \rowcolor{gray!25}
  DeepCity & 0.48 & 0.63 & 0.44 & 0.59 & 0.41 & 0.54 \\
  DeepCity$(\usr)$ & 0.42 & 0.58 & 0.42 & 0.56 & 0.40 & 0.54 \\
  Tensor & 0.25 & 0.51 & 0.22 & 0.50 & 0.27 & 0.43 \\
  MF & 0.25 & 0.50 & 0.21 & 0.50 & 0.27 & 0.43 \\
  SAP & 0.29 & 0.55 & 0.31 & 0.56 & 0.20 & 0.38 \\
  DeepWalk & 0.33 & 0.51 & 0.33 & 0.50 & 0.27 & 0.42 \\
  \bottomrule[\thickline]
  \end{tabular}
  \caption{Experimental results on location category prediction.\label{table:predict_cat}}
\end{table*}

Notice that all the parameter in the baseline models are set
according to the original settings presented in their papers.
We feed the same set of users and locations into DeepCity 
and the baseline models,
in order to eliminate the possible sample bias 
introduced to the prediction result.

We adopt logistic regression as the learning algorithm:
user gender is predicted through a binary classifier 
while the other three tasks 
are predicted through one-vs-rest logistic regression.
We randomly split the learned features 
with 70\% for training classifiers
while the left 30\% for testing.
The random split is repeated for 10 times
and we report the average results.
We adopt AUC (area under the ROC curve), Precision, Recall and F1 score
as metrics for gender prediction
while Macro-F1 and Micro-F1 for age, race and location category prediction.

There are several parameters in DeepCity,
as shown in Algorithm~\ref{alg:gen},
including the length of each walk ($s$),
number of walks per node ($r$), 
dimension of learned features ($d$).
We follow the settings 
of DeepWalk and node2vec,
i.e., $s=80$, $r=10$ and $d=128$.

\subsection{Results}
The experimental results of four prediction tasks are presented 
in Tables~\ref{table:predict_gender},~\ref{table:predict_race},~\ref{table:predict_age}
and~\ref{table:predict_cat}.
From the results,
we observe that DeepCity
for demographic and location category prediction
achieves a superior performance,
e.g., the AUC score for gender prediction 
is 0.95 for all cities
and more than 0.8 precision and recall are achieved at the same time,
and it outperforms all the baselines across all prediction tasks.
This indicates that 
our DeepCity framework
is promising on mining user check-ins.
Moreover, 
DeepCity outperforming DeepCity$(\loc)$
demonstrates the importance of temporal information
in predicting user demographics.
The same conclusion can be drawn 
from location category prediction (see Table~\ref{table:predict_cat}).
Meanwhile, DeepCity$(\cathour)$ and DeepCity$(\cat)$ achieving
a worse performance than DeepCity
suggests that location category, 
even at a fine-grained level 
(the second layer of Foursquare's category tree),
is not enough for profiling users' demographics.
Alternatively, from user privacy point of view,
this suggests that replacing detailed location information sharing
with location categories (semantic locations)
can help users to gain better privacy protection.

The reason why STL does not perform well in our experiments
might be caused by the data sparseness.
Also, as mentioned previously, 
the lack of locations' reviews and keywords
may result in the relatively low performance,
since reviews and keywords 
contain resourceful information of each location
which can be used to describe users' demographics.
The matrix factorization approach,
on the other hand, performs the worst as 
it does not take into account enough information.
For location category prediction,
DeepCity outperforms SAP,
one reason might be that SAP's information retrieval feature set
considers only one hop neighbors of each location,
and the connections established by users in SAP
can introduce many noises on grouping locations of the same category together.
On the other hand,
our method takes into account more than one hop neighbors
of each location, i.e., the context size in Skip-gram~\cite{MCCD13,MSCCD13},
and the location category biased values of temporal-users
mitigate the noises introduced
on location connections.

We also observe that DeepCity
outperforms DeepWalk in all prediction tasks,
this indicates that task-specific random walk
indeed increases the prediction performance
over general network embedding methods,
which validates our intuitions presented previously.
This result should not only
be limited to mining user check-ins,
but also other data mining fields,
such as profiling users through published statues in OSNs.
We leave the further investigation as one future work.
However, even though without any adjustments,
DeepWalk still achieves competitive performances,
which further suggests network embedding's effectiveness
in prediction tasks.

\subsection{Parameter sensitivity}

DeepCity involves a number of parameters,
we concentrate on the sensitivity of three of them,
including the length of each walk ($s$), number of walks per node ($r$), 
and dimension of learned features ($d$).
Except for the parameter being tested,
others are set to their default values.

\begin{figure*}[!t]
  \centering
  \begin{minipage}{0.67\columnwidth}
  \includegraphics[width=\columnwidth]{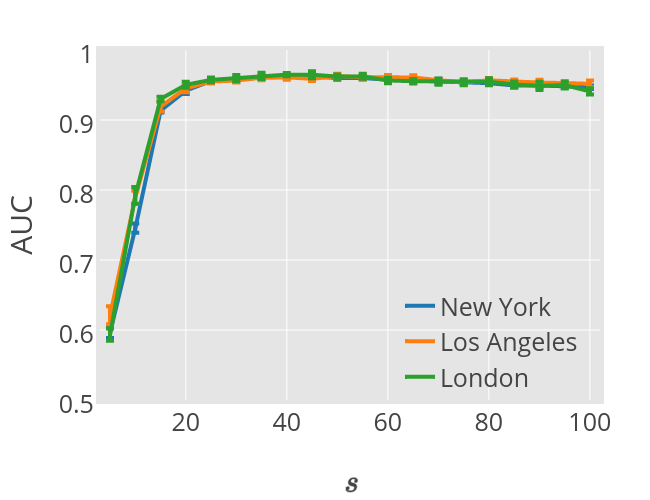}
  \end{minipage}
  \begin{minipage}{0.67\columnwidth}
  \includegraphics[width=\columnwidth]{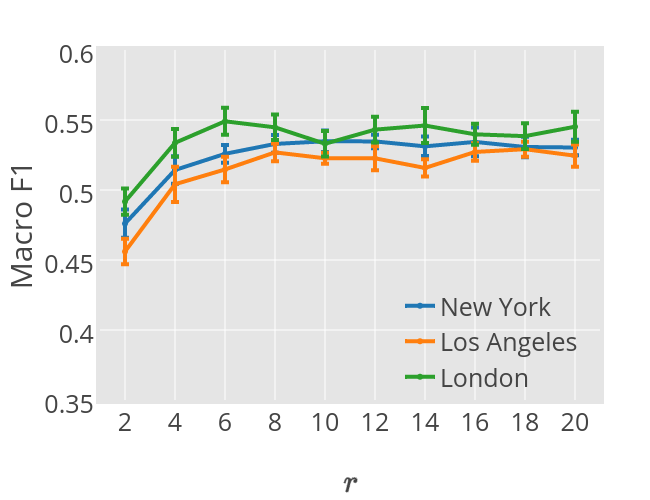}
  \end{minipage}
  \begin{minipage}{0.67\columnwidth}
  \includegraphics[width=\columnwidth]{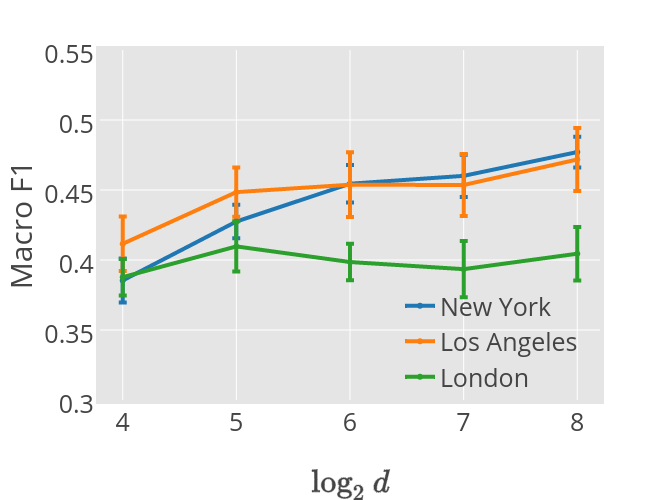}
  \end{minipage}
  \caption{Parameter sensitivity study with 
  gender (left),  age (middle) and category (right) prediction.\label{fig:para}}
\end{figure*}

We utilize gender prediction to study the sensitivity of $s$.
Figure~\ref{fig:para} (left)
shows that the prediction result (AUC)
experiences a dramatic increase when $s$ changing
from 5 to 20, and then stay stable.
Similarly, the number of walks starting from each node
increases the performance of age prediction (Macro-F1)
when shifted from 2 to 8 and saturate then (Figure~\ref{fig:para} (middle)).
The reason that the performance gain by increasing $r$ 
is not that large 
compared with $s$
is that $s=80$ in the default setting 
already provides Skip-gram with a large quantity data from embedding.
The similar situation can be observed from varying $d$ (Figure~\ref{fig:para} (right)),
especially in London, the Macro-F1 on location category prediction
is rather stable.
We also study other parameters' sensitivity,
such as the context window size used 
in the Skip-gram model (its default value is 10 following DeepWalk and node2vec), 
the prediction results do not vary much
and are omitted here.

\begin{figure*}[!t]
  \centering
  \begin{minipage}{0.67
  \columnwidth}
  \includegraphics[width=\columnwidth]{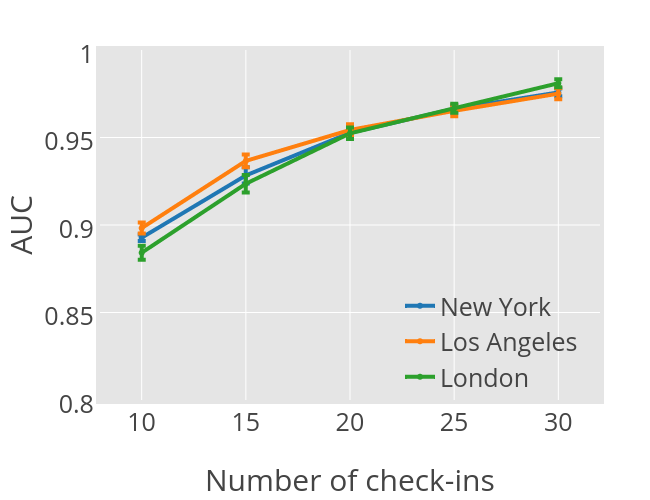}
  \end{minipage}
  \begin{minipage}{0.67\columnwidth}
  \includegraphics[width=\columnwidth]{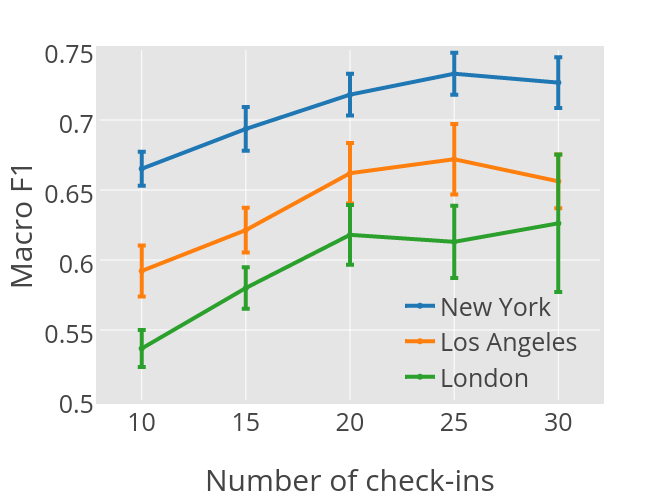}
  \end{minipage}
  \begin{minipage}{0.67\columnwidth}
  \includegraphics[width=\columnwidth]{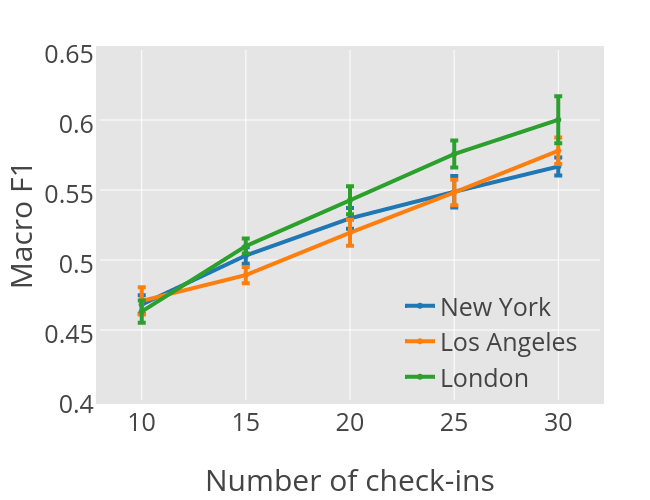}
  \end{minipage}
  \caption{Robustness study
  with gender (left), race (middle) and age (right) prediction.\label{fig:robust}}
\end{figure*}


\subsection{Robustness study}
As mentioned previously,
our demographic prediction is based on users with at least 20 check-ins,
i.e., active users.
However, how to define active users,
i.e., with at least how many check-ins, is not clear.
In the literature,
various choices have been made, such as 10 in~\cite{CML11},
2 in~\cite{GTHL15}, 
1,000 in~\cite{CCL10},
and 40 in~\cite{WL16}.
Therefore, 
we further study the different choices of defining active users
with respect to prediction performance.

Figure~\ref{fig:robust}
depicts the prediction results
as a function of the least number of check-ins for each active users.
For gender prediction, 
the AUC grows almost linearly 
as the increase of check-ins.
The similar situation is observed in age prediction.
Meanwhile, the result of race prediction arrives at a stable level
after 25 check-ins.
In general, the results are not surprising since more check-ins a user shares,
more information we obtain from him,
higher chance we can predict his demographics correctly.
We further notice that
when the least number of check-ins for each user 
is set to 10,
we still achieve a very strong performance,
e.g., AUC is approaching 0.9 for gender prediction,
this further shows the effectiveness of DeepCity
on profiling users' demographics.
We also study the robustness of location category prediction
as a function of the least number of check-ins at each location,
the results are similar and are not shown.

\section{Related Work}
\label{sec:relwork}
\noindent\textbf{Network embedding.}
Network embedding based on deep learning techniques 
has received many attentions during the past three years.
Perozzi et al.~\cite{PAS14} are among the first
to utilize the neural network model, Skip-gram,
for network embedding.
The basic analogy of DeepWalk is:
each node is treated as a word 
and random walk generated node sequences are treated as sentences.
The authors of~\cite{PAS14} have performed a multi-label classification task
with DeepWalk's learned features,
experimental results show its superior performances 
over traditional network embedding methods, such as spectral clustering.
Inspired by DeepWalk, the authors of~\cite{TQWZYM15} 
propose another network embedding method, namely LINE.
The object function of LINE is designed 
to preserve both the global and local network structures,
which are complementary to each other.
To increase the training efficiency,
the authors of~\cite{TQWZYM15} 
further propose an edge sampling algorithm.
Experiments on various networks have demonstrated LINE's effectiveness
in both prediction and visualization.
More recently,
Grover and Leskovec propose an embedding method, namely node2vec~\cite{GL16}.
Similar to DeepWalk,
node2vec utilizes random walks 
to simulate sentences which later are feed into the Skip-gram model
to obtain embedding vectors.
However, node2vec proposes a flexible notion of a node's neighbors
by introducing several tunable parameters in the random walk,
with which a node's diverse neighborhood can be explored.
In addition to node embedding, 
node2vec also introduces edge embedding
which is based on
multiple binary operations between two node's vectors.
Experiments on multi-label classification and link prediction
have shown node2vec's effectiveness.
All the above methods' aim
is to learn nodes' general features
for performing multiple predictions,
DeepCity, on the other hand, utilizes task-specific random walk 
to produce different features for different predictions.
Experimental results have shown that 
the features learned by DeepCity indeed outperforms general purpose ones, i.e., DeepWalk,
which validates our intuition in Section~\ref{sec:framework}.

\smallskip
\noindent\textbf{Mining user check-ins.}
With the large amount of user check-in data being available,
researchers have concentrated on mining these data.
One direction is to use user check-ins to predict friendships,
such as~\cite{CBCSHK10,SNM11}.
Meanwhile,
a few works have explored a user's friends information
to predict his locations~\cite{BSM10,CML11}.
Recently, check-in data are used to profile users,
e.g, the STL model~\cite{ZYZZX15} 
adopted by us as a baseline for demographic prediction.
On the other hand,
many works have been conducted on reshaping
our understandings of locations from the user aspects,
such as the SAP model~\cite{YSLYJ11}.
Researchers have also used check-in data
to measure the happiness~\cite{QSA14} 
and walkability~\cite{QASD15} of streets,
and find the similar neighborhoods across different cities~\cite{FGM15}.

\section{Conclusion}
\label{sec:conclu}

In this paper, we propose a general framework, namely DeepCity,
to learn features for user and location profiling.
We propose a method, i.e., task-specific random walk,
to guide the learning algorithms 
to embed users with similar demographics
(locations with similar categories)
closer in the resulted vector space.
Experimental results on a large collection of Instagram data
have demonstrated the effectiveness of DeepCity
over other models.

There are several directions we plan to pursue
in the future.
First, as mentioned in Section~\ref{sec:relwork},
many works have concentrated on the mutual prediction
between social network and check-ins.
We plan to apply DeepCity 
to tackle these problems as well.
Second, as stated in Section~\ref{sec:experiments},
the task-specific random walk is not limited 
to mining user check-ins,
we aim to apply our framework to 
profile users with other information sources,
such as Tweets.

\bibliography{dm}
\bibliographystyle{abbrv}

\end{document}